\documentclass{osa-article}
\usepackage{amsmath}
\journal{oe}

\articletype{Research Article}

\begin{document}

\title{Ghost Imaging with the Optimal Binary Sampling}

\author{Dongyue Yang\authormark{1}, Guohua Wu\authormark{1,*}, Bin Luo\authormark{2} and Longfei Yin\authormark{1} }

\address{\authormark{1}School of Electronic Engineering, Beijing University of Posts and Telecommunications, Beijing, 100876, China\\
\authormark{2}State Key Laboratory of Information Photonics and Optical Communications, Beijing University of Posts and Telecommunications, Beijing, 100876, China\\
}

\email{wuguohua@bupt.edu.cn} 


\begin{abstract}
To extract the maximum information about the object from a series of binary samples in ghost imaging applications, we propose and demonstrate a framework for optimizing the performance of ghost imaging with binary sampling to approach the results without binarization. The method is based on maximizing the information content of the signal arm detection, by formulating and solving the appropriate parameter estimation problem - finding the binarization threshold that would yield the reconstructed image with optimal Fisher information properties. Applying the 1-bit quantized Poisson statistics to a ghost-imaging model with pseudo-thermal light, we derive the fundamental limit, i.e., the Cram{\'e}r-Rao lower bound, as the benchmark for the evaluation of the accuracy of the estimator. Our theoertical model and experimental results suggest that, with the optimal binarization threshold, coincident with the statistical mean of all bucket samples, and large number of measurements, the performance of binary sampling GI can approach that of the ordinary one without binarization.
\end{abstract}

\section{Introduction}
\newcommand{\RNum}[1]{\uppercase\expandafter{\romannumeral #1\relax}}
\indent Ghost imaging (GI), which used to be considered as a counter-intuitive phenomenon from the first time it was demonstrated \cite{pittman1995optical}, allows an unknown object to be obtained by measuring the spatial-temporal properties of a light beam that never interacts with it. This indirect imaging method relies on the intensity or fluctuation correlation between a point-like bucket detection and a spatial-resolved non-contact reference profile, either a light field detection \cite{valencia2005two,Xue2014Lensless}, or a pattern modulation \cite{bromberg2009ghost}. Compromisingly, a large number of repeated measurements are required for reconstructing a high quality image \cite{Sun2012Normalized,li_negative_2017,Wang:16}, which has become a major drawback preventing GI from practical applications, especially real-time tasks, even with the help of compressive sensing technique \cite{katz2009compressive}. Considering this, reducing the dynamic range of detectors or recording measurements with less bits, even 1-bit, would speed up GI process significantly when there are less data to be sampled, transported, stored, and calculated \cite{li_image_2017}. In fact, it is even more suitable for computational GI \cite{shapiro2008computational}, where the reference camera is replaced by a spatial light modulator, thus the sampling process of reference camera is equivalent to the pattern modulation of spatial light modulator. Binary modulation would bring a much higher modulation rate, especially for the digital micro-mirror device (DMD) \cite{duarte2008single}, which is in nature a two-level device and has to conduct period multiplexing to accomplish gray scale modulation.\\
\indent For the GI applications under extreme weak echo response conditions \cite{liu2018fast,li_single-photon_2019,Deng:17}, i.e., detectors hardly register more than one photon for each measurement, the detection noise would impose a significantly negative impact on the image reconstruction of GI \cite{shi2018image}. Binary sampling has provided an effective way to suppress the background noise at low-light-level \cite{ke2016fast}, also capable of high sensitivity, which is the inherent binary nature of single-photon avalanche detectors (SPAD), one of the most sensitive device to measure the light intensity. This natural combination of binary sampling and SPAD could benefit GI in both sensitivity and robustness against noise.\\ 
\indent However, binary sampling have its drawbacks in the loss of information. When we digitalize the signal into 1-bit, we create at each output a quantization error: the difference between the original signal and the binarization threshold. This quantization error does harm the image quality of GI \cite{li_image_2017}. Here comes the question - given a binary sampling GI scenario with certain characteristics (e.g. laser light intensity, number of camera pixels, measurement number) - how good can this binary sampling GI perform? Or given a goal of image quality - how can we optimize the characteristics of our GI system? Is there an optimal binary detection that would yield maximal physical information about the object? To answer these questions, we treat the GI procedure as an estimation of the coincidence measurements for decoding the brightness distribution of the object. A powerful measure of the effectiveness of this procedure is based on Fisher information \cite{frieden1972restoring,kay1993fundamentals}, a concept from statistical information theory. Fisher information is a mathematical measure of the sensitivity of an observable quantity (e.g. image quality assessments) to changes in its underlying parameters (e.g. binarization threshold). Using the Fisher information function, one may compute the Cram{\'e}r-Rao lower bound (CRLB), which provides a theoretical lower bound on the variance of an unbiased estimator. With the right estimator, the performance of GI, represented by the CRLB, can be evaluated quantitively.\\
\indent The purpose of this work is to fundamentally investigate the binary sampling in GI. We propose a framework for optimizing the performance of binary sampling ghost imaging. Our image estimation model based on the measure of Fisher information reveals an informative-optimal binarization threshold for the samples of the signal arm, which is the statistical mean of all bucket samples. With the optimal binarization threshold and large measurement number, the performance of binary sampling GI can approach the ordinary one without binarization. The results of the designed experiments demonstrate highly agreement with the predictions of our model.\\
\section{Theoretical Analysis}
\subsection{Equivalent imaging model of GI}
In classical imaging, the image sought could be considered as a parametric approximation of $\lambda(x)$ \cite{thiebaut2017principles}, the (normalized) object brightness distribution, written as\\
\begin{equation}
\lambda(x)= \sum_{n=1}^{N} s_nh(Nx-n),\label{5}
\end{equation}
where $s_n$ represents the reflective or transmissive function of the object, with $n$ being the number of subfields on the object plane ($1\le n < N$), $h(x)$ is the point spread function (PSF) on the image plane. Due to the finite size of the lens, the impulse response cannot be a Dirac delta, which builds a "point-to-spot" correspondence from a geometrical light point on the object plane to a unique geometrical light spot on the image plane, inducing a resolution limit of the imaging system. Similarly, the physics of second-order-measuring GI diagram can also be understood via Klyshko's "unfolded picture" \cite{Klyshko1988The,Belinskii1994Two}, which also build a "point-to-spot" correspondence to make the "ghost" image of the object aperture possible \cite{shih2014introduction}. \\
\indent In the equivalent imaging model of GI, as shown in Fig. 1, $s_{n,j}$ refers to a subfield of the diffuser located on spatial unit $n$ at $j$th measurement, corresponding to the spatial unit $x$ filtered by the object $O(x)$, and the light emitted from the source is then divided by a beam splitter (BS) into the reference and the signal arm, where GI requires the same light intensity distribution $\lambda_{j}(x)$ on both the object plane and reference plane in each $j$th measurement. To simplify the presentation, we base our discussions on a 1D sensor array, but all the results can be easily extended to 2D case.\\
\subsection{Reference detection}
\begin{figure}[htbp]
\centering
\includegraphics[width=1\textwidth]{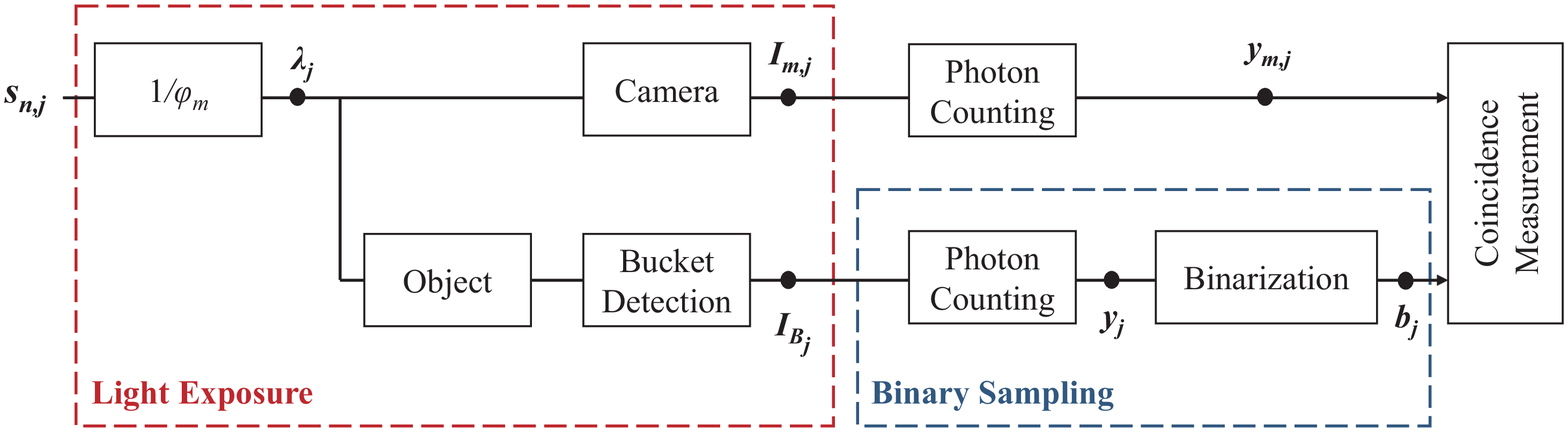}
\caption{Signal processing diagram of the equivalent imaging model of binary sampling GI.}
\label{Fig1}
\end{figure}
\indent On the reference plane, a spatial-resolved image sensor, i.e., a camera works as a sampling device of $\lambda_{j}(x)$. Suppose that the sensor consists of $M$ pixels on the area of interest (AOI), assumed to be the same scale of $O(x)$, the $m$th pixel covers an area between $\left[ m/M,m+1/M \right)$, for $m\in[1,M)$. Besides, GI requires $J$ independent coincidence measurements to reconstruct an image. For the $j$th ($\in[1,J]$) frame or modulation pattern, within an exposure time $\tau$, we denote the light intensity accumulated on the $m$th pixel by $I_{m,j}$, 
\begin{equation}
\begin{aligned}
I_{m,j} & \stackrel{\text { def }}{=} \int_{0}^{\tau} \int_{m / M}^{(m+1) / M} \lambda_{j}(x) d x d t \\ &=\tau\langle\lambda_{j}(x), \beta(M x-m)\rangle,
\end{aligned}
\end{equation}
where $\beta(x)$ is a box function, reads,
\begin{equation}
\beta(x) \stackrel{\text { def }}{=}\left\{\begin{array}{ll}{1,} & {\text { if } 0 \leq x \leq 1;} \\ {0,} & {\text { otherwise. }}\end{array}\right.
\end{equation}
\indent Substituting Eqs. (1) and (3) into (2), we have
\begin{equation}
I_{m,j} =\frac{\tau}{N} \sum_{n} s_{n,j}\left\langle h(x), \beta\left(\frac{M(x+n)}{N}-m\right)\right\rangle,
\end{equation}
\indent ${I_{m,j}}$ denote the exposure values accumulated by the $m$th sensor pixel in $j$th measurement, which has a stochastic relation to $y_{m,j}$, the number of photons impinging on the surface of the $m$th pixel during the exposure time $\tau$ in $j$th measurement. More specifically, according to the semi-classical theory of photoelectric detection \cite{goodman2015statistical}, $y_{m,j}$ can be modeled as realizations of a Poisson random variable $Y_{m,j}$ \cite{helstrom1969quantum}, with intensity parameter ${I_{m,j}}$, i.e.,\\
\begin{equation}
\mathbb{P}\left(Y_{m,j}=y_{m,j} ; I_{m,j}\right)=\frac{I_{m,j}^{y_{m,j}} e^{-I_{m,j}}}{y_{m,j} !}, \quad \text { for } y_{m,j} \in \mathbb{Z}^{+} \cup\{0\}
\end{equation}
\indent The expectation (statistical mean) of this Poisson process is $\mathbb{E} [Y_{m,j} ]=I_{m,j}$, since the average number of photons captured by a given pixel is equal to the local light exposure $I_{m,j}$. Considering the random nature of the $s_{n,j}$, both in spatial $(n)$ and time $(j)$ domain, caused by the pseudo-thermal property \cite{goodman2015statistical} and independent repeated measurement in GI, we simplify the marginal distribution of $s_{n} = \sum_{j} s_{n,j} $, $\lambda(x)= \sum_{j}\lambda_{j}(x)  $, and $ I_{m} = \sum_{j} I_{m,j}$.\\
\subsection{Bucket detection \& Binary sampling}
\indent For modeling of the "point-like" bucket detection in the signal arm, the bucket detector can be considered as a collection of pixels filtered by the object $O(x)$ with an intensity summation output. In what follows, and without loss of generality, we assume the $O(x)$ is box function with a length of $M$, which can be imagined to be a $M$ length "single-slit" object. Recall that the light field $\lambda(x)$ on the object plane, and the number of photons impinging on each pixel of the spatial unit subject to the Possion process with a same parameter $I_{m}$, according to the central limit theorem (CLT) when the number $M$ of pixels is large, the bucket signals are $\{I_{B_{j}} = MI_{m,j}\}$, and the photon counts subject to a Gaussian random variable $\left\{Y_{B_{j}}\right\} \sim N(MI_{m},MI_{m}) $ in time $(j)$ domain. Using the approximation of Gaussian to Poisson process with a large $MI_{m}$, we denote $\mathbb{P}(Y_{B_{j}}=y_{j}; MI_{m}) = p(MI_{m})$, i.e.,
\begin{equation}
\mathbb{P}\left(Y_{B_{j}}=y_{j}; MI_{m}\right)=\frac{(MI_{m})^{y_{j}} e^{-MI_{m}}}{ y_{j} !}. \quad \text { for } y_{j} \in \mathbb{Z}^{+} \cup\{0\}
\end{equation}
\indent In this contribution, we apply binarization only on the bucket signals. However, this binarization optimization framework can be easily extended to the samples of reference detectors, but not included here for simplicity. The binary output ${b_{j}}$ are drawn from the mapping of random variables $Q:Y_{B_{j}} \longmapsto B_{j}$, such that\\
\begin{equation}
Q(y)=\left\{\begin{array}{ll}{1,} & {\text { if } y \geq q;} \\ {0,} & {\text { otherwise. }}\end{array}\right.
\end{equation}
where $q>0$ is the binarization threshold. Introducing the probability distribution function (PDF) as $\mathbb{P}(B_{j}=b_{j} ; MI_{m})=p_{b_{j}}\left(MI_{m}\right), \ b_{j} \in\{0,1\}$, i.e.,
\begin{eqnarray} 
&& p_{0}(MI_{m}) \stackrel{\text { def }}{=} \sum_{k=1}^{q} \frac{(MI_{m})^{k}}{k !} e^{-MI_{m}}, \\
&& p_{1}(MI_{m}) \stackrel{\text { def }}{=} 1- p_{0}(MI_{m}) = 1-\sum_{k=1}^{q} \frac{(MI_{m})^{k}}{k !} e^{-MI_{m}},
\end{eqnarray} 
\subsection{Performance Analysis}
\indent The previous discussions reveal the relation between the subfield $s_{n,j}$ and the double arm coincident measurements at $j$th frame, and GI relies $J$ consecutive repeated measurements to reconstruct the image $T(m)$ of object by the means of second-order correlation function, 
\begin{equation}
T(m) = \frac{\left\langle Y_{m,j}  \times B_{j} \right\rangle_{j}}{\langle Y_{m,j}\rangle_{j}\langle B_{j}\rangle_{j}},
\end{equation}
where $\langle\cdot\rangle_{j}$ denotes average overall the measurements, and $Y_{m,j}$, $B_{j}$ (or $Y_{B_{j}}$, without binarization) represent the intensity distribution at the reference and binarized bucket detection, respectively. After modeling the detection process, reconstructing the image of object boils down to estimating the unknown deterministic parameters $\{s_{n}\}$. Input of our estimation problem is two coincidence sequences of binary samples of bucket detector $\{B_{j}\}$ and ideal samples of the camera $\{Y_{m,j}\}$. The PDF of $\{B_{j}\}$ depends on the averaged light intensity $\{I_{m}\}$ over all measurements, which are linked to the subfield parameter $\{s_{n}\}$ of the light source. In our analysis, we assume that the GI system is piecewise stable, i.e., $M$, $N$ are constant and the PSF $h(x)$ in Eq. (1) has a permanent distribution, the spatial sampling factor $1/\varphi_{m}$ is a constant supported within $\left[1,M/N\right]$ \cite{yang2011bits}. Noting that convolution can be considered as a linear operator, the mapping between $s_{n}$ and $I_{m}$ can now be simplified as 
\begin{equation}
I_{m} = s_{n}/\varphi_{m} ,\quad for \ \ \ m\in\left[\frac{M(n-1)}{N}+1, \frac{Mn}{N}\right]
\end{equation}
\indent Besides, due to the random modulation of R.G.G. and pseudo-thermal light property, the coincidence measurements are assumed to be independent and identical distributed (iid.) on the time domain, i.e., the temporal sampling  factor $1/\omega_{j}$ could also be considered as a constant, which simplifies the mapping between $s_{n,j}$ and $I_{m,j}$, reads,
\begin{equation}
I_{m,j} = s_{n,j}/\varphi_{m}\omega_{j} ,\quad for \ \ \ m\in\left[\frac{M(n-1)}{N}+1, \frac{Mn}{N}\right],\ j\in [1,J]
\end{equation}
\indent It is apparent that the parameters $\{s_{n}\}$ have disjoint spatial regions of influence, thus we can do the estimation one-by-one, independently of each other. Without loss of generality, we focus on the estimation of parameter $s_1$ from a sequence $b_{j} = \left[ b_{1},\ldots,b_{J}\right]^{T}$ and  $I_{m,j} = \left[ I_{1,1},\ldots,I_{M/N,J}\right]^{T}$. For simplicity, we drop the subscript of $s_1$ and use $s$ instead. The likelihood function $\mathcal{L}_{\boldsymbol{b}}(s)$ of observing the binary sampling coincidence measurements is written as,
\begin{equation}
\begin{aligned} 
\mathcal{L}_{\boldsymbol{b}}(s) & \stackrel{\text { def }}{=} \mathbb{P}\left(B_{j}=b_{j}, j\in[1,J] ; s\right) \mathbb{P}\left(Y_{m,j}=y_{m,j},m\in[1,M/N), j\in[1,J] ; s\right) \\
&=\prod_{j=1}^{J} p_{b_{j}}(Ms/\varphi_{m}) \prod_{j=1}^{J}\prod_{m=1}^{M/N} p(s/\varphi_{m}\omega_{j}).
\end{aligned}
\end{equation}
\indent Defining $J_{1}$ ($\in [1,J]$) to be the number of "1"s in the binary sequence, we can simplify Eq. (13) as
\begin{equation}
\mathcal{L}_{b}(s)=\left(p_{1}(Ms /\varphi_{m})\right)^{J_{1}}\left(p_{0}(Ms / \varphi_{m})\right)^{J-J_{1}} \prod_{j=1}^{J}\prod_{m=1}^{M/N} p(s/\varphi_{m}\omega_{j}),
\end{equation}
\indent In order to measure the sensitivity of a measurement (in our case, the coincidence measurement) to the parameters being estimated (source field $s$), we introduce the Fisher information matrix and the CRLB \cite{kay1993fundamentals}, since the accuracy of parameter $s$, its true value, is at best equal to the square root of the CRLB \cite{chao_fisher_2016}. For the Fisher information matrix in our case, which can be simply written as $ I(s)=\mathbb{E}\left[-\frac{\partial^{2}}{\partial s^{2}} \log \mathcal{L}_{b}(s)\right]$. By substituting Eq.(14) and after some straight manipulations, $I_{b}(s)$ can be simplified as\\
\begin{equation}
\begin{aligned} 
I_{b}(s)&=\mathbb{E}\left[-\frac{\partial^{2}}{\partial s^{2}}\left(J_{1} \log p_{1}(Ms /\varphi_{m})+\left(J-J_{1}\right) \log p_{0}(Ms / \varphi_{m})+\sum_{j=1}^{J} \sum_{m=1}^{M/N} \log p(s /\varphi_{m}\omega_{j})\right)\right]\\
&=\frac{JM^{2}}{\varphi_{m}^{2}}\frac{p_{0}^{\prime}(Ms /\varphi_{m})^{2}}{p_{0}(Ms /\varphi_{m}) p_{1}(Ms /\varphi_{m})} +\mathbb{E}\left[\sum_{j=1}^{J} \sum_{m=1}^{M/N}\left(y_{m, j}\right)\right] / s^{2},
\end{aligned} 
\end{equation}
\indent Using the definition of $p_{0}(x)$ in Eq. (8), the derivative $p_{0}^{\prime}(x)$ can be computed as
\begin{equation}
p_{0}^{\prime}(x)=-e^{-x} \frac{x^{q}}{(q) !}.
\end{equation}
\indent Since $\{y_{m,j}\}$ are drawn from Poisson distributions as in Eq. (5), we have $\mathbb{E}[y_{m,j}] = I_{m,j} = s/\varphi_{m}\omega_{j} $ for all $m$. Then,
\begin{equation}
\mathbb{E}\left[\sum_{j=1}^{J} \sum_{m=1}^{M/N}\left(y_{m, j}\right)\right] / s^{2}  =\frac{JM}{N}( \frac{s}{\varphi_{m}\omega_{j}})\frac{1}{s^{2}}=\frac{JM}{N\varphi_{m}\omega_{j}}\frac{1}{s},
\end{equation}
\indent Substituting Eqs. (7), (15) and (16) into (14), using the definition of $CRLB_{b} = 1/I_{b}(s)$, and after some straightforward manipulations, we have
\begin{equation}
\begin{aligned} 
CRLB_{b} &= \frac{s}{JM}\frac{N\varphi_{m}\omega_{j}\Gamma}{N\omega_{j}+ \Gamma},\\
where \ \ \Gamma &= \sum\limits_{j=1}^{q} \frac{(q) !(Ms /\varphi_{m})^{-j}}{(q-j) !}\sum\limits_{j=1}^{\infty} \frac{(q) !(Ms /\varphi_{m})^{j}}{(q+1+j) !}.
\end{aligned} 
\end{equation}
\indent For comparison, the case without any binarization are also investigated, where bucket outputs are $y \stackrel{\text { def }}{=}\left[y_{1}, y_{2}, \ldots, y_{J}\right]^{T}$, i.e., the number of photons impinging on each pixel. The likelihood function $\mathcal{L}_{\boldsymbol{y}}(s)$ of the ideal coincidence measurement in this ideal case is,\\
\begin{equation}
\begin{aligned} 
\mathcal{L}_{\boldsymbol{y}}(s) & \stackrel{\text { def }}{=} \mathbb{P}\left(Y_{j}=y_{j}, j\in[1,J]; s\right) \mathbb{P}\left(Y_{m,j}=y_{m,j},m\in[1,M/N), j\in[1,J] ; s\right) \\ 
&=\prod_{j=1}^{J} p(s/\varphi_{m}) \prod_{j=1}^{J}\prod_{m=1}^{M/N} p(s/\varphi_{m}\omega_{j}), 
\end{aligned}
\end{equation}
\indent Using the Fisher information $I_{i}(s) = \mathbb{E}\left[-\frac{\partial^{2}}{\partial s^{2}} \log \mathcal{L}_{y}(s)\right]$ and similar calculation process of Eq. (17), we get \\
\begin{equation}
\begin{aligned} 
CRLB_{i}  = \frac{s^{2}}{\mathbb{E}\left[\sum\limits_{j=1}^{J}\left(y_{j}\right)\right] + \mathbb{E}\left[\sum\limits_{j=1}^{J} \sum\limits_{m=1}^{M/N}\left(y_{m, j}\right)\right]} = \frac{s}{JM}\frac{N\varphi_{m}\omega_{j}}{N\omega_{j}+1},
\end{aligned}
\end{equation}
\begin{figure}[htbp]
\centering
\includegraphics[width=1\textwidth]{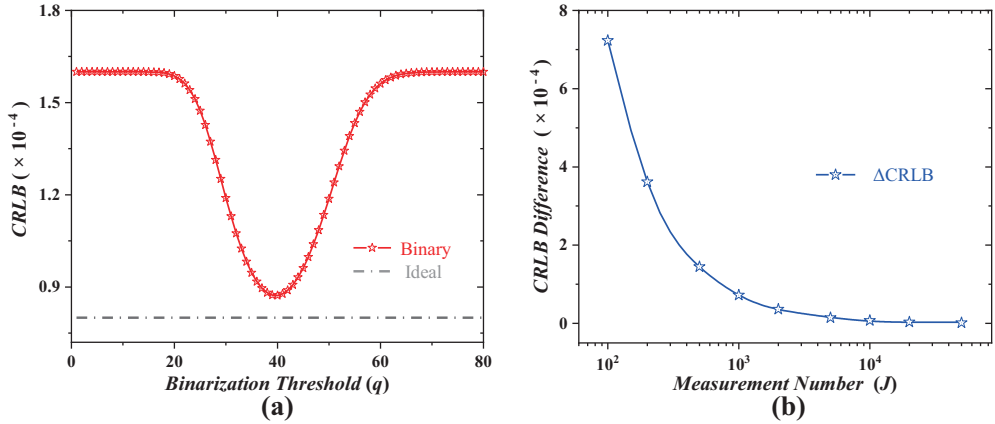}
\caption{Simulation results: (a) The CRLB of binary sampling GI vs. $q$, compared with the CRLB of ideal sampling GI. Source intensity $s=8$, number of measurement $J=10,000$, pixels $M=5$, subfields $N=1$ are considered. (b) The difference of CRLB vs. measurement number under the optimal $q=Ms/\varphi_{m}=40$. Data fitted to an exponential decay function. }
\label{Exp1}
\end{figure}
\indent For the comparison between the performance of binary sampling GI under different $q$, we do the numerical simulations to calculate the CRLB of different schemes with certain parameters. For reducing the computational complexity, the spatial sampling factor are assumed to be $\varphi_{m}=1$, refers to a PSF of box function when $m\in[1,M/N]$, and the temporal sampling factor $\omega_{j}=1$, refers to a uniform sampling when $j\in[1,J]$. The behavior of CRLB of different sampling schemes against threshold $q$ is shown in Fig.2 (a). With the optimal $q=Ms/\varphi_{m}$, CRLB of binary sampling GI reaches the minimum, indicating the optimal estimation of $s$, yields the highest quality of reconstructed image of binary GI. The physical meaning of this optimal threshold can be simply understood that the temporally-averaged light field $s/\varphi_{m}$ emitted from the subfield $s$ is filtered by a M-pixles object and summed by a bucket, which is the statistical mean of bucket outputs. Besides, the difference between the two CRLBs could also provide a measure of performance degradations incurred by the binary sampling operation, as shown in Fig. 2(b). Under the optimal $q$, the best-precision gap between the binary sampling GI and the ordinary one, denoted by the difference $\Delta CRLB = CRLB_{b}-CRLB_{i}$, are subjected to a negative exponential decay behavior and converging to 0 with the increasing measurement number $J$, which heralds the clue of the reconstructed image of binary sampling GI approaching the ordinary one under large number of measurements.\\

\section{Experimental Verification}
\begin{figure}[htbp]
\centering
\includegraphics[width=0.7\textwidth]{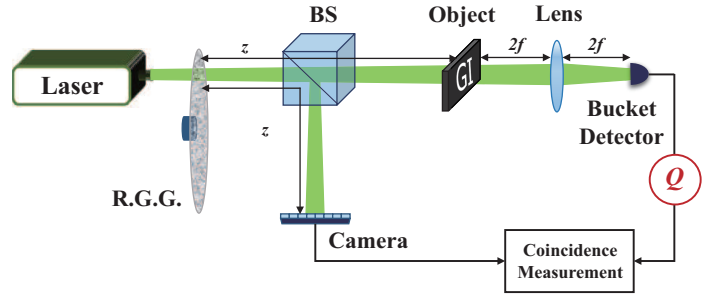}
\caption{Experimental setup. Red mark $Q$: binarization. R.G.G., rotating ground glass. BS, beam splitter.}
\label{Figure3}
\end{figure}
\indent Experiments are designed to verify the above analysis of the performance of binary sampling GI. For the conventional GI setup \cite{gatti2004ghost,Nan:17}, as shown in Fig. 3. The pseudo-thermal light source, which mainly consists of a 532 nm CW laser and a rotating ground glass (R.G.G.), generates random speckle patterns. The intensity of the laser is 5 mW, the rotating speed of R.G.G. is 0.32 rad/s, $z$ = 300 mm, $f$ = 100 mm. The light emitted from the source is then divided by a beam splitter (BS) into the reference and the signal arms. The signal arm penetrates a transmissive object aperture, a 'GI' pattern of 500 $\mu$m, be focused by lens and then to be registered by a point-like bucket detector (Thorlabs PDA100A2) as an intensity sequence. The spatial profile, $I_{m,j}$, comes from the reference arm, which never interact with the object, is recorded by a commercial CMOS camera (XiMEA MQ003CM) with an AOI of $140\times 140$ pixels, placed on the image plane and synchronically triggered with the bucket detector. The image is reconstructed by the second-order correlation defined in Eq. (10). \\
\indent We introduce binarization on the outputs of the bucket detector to mimic the binary detection, as shown in Fig. 3. Binarization with different threshold $q$ is applied to the recorded quasi-continuous signals of each bucket measurement, as the mapping $Q$ defined  in Eq. (7). Here we want to mention that, although this can easily extend to the reference samples, it is not included in our case for simplicity and practical considerations, since the reference arm of some GI applications usually be compacted into the transmitting system with the light source, as described in Ref.\cite{wang_airborne_2018, zhao2012ghost, Li:15}, so the sampling bits of reference detectors would not be compressed in our experiment.\\
\indent When we change the binarizaiton threshold $q$ on the bucket signals, the best and worst reconstructed images under 20,000 measurements are shown in Fig. 4, row \RNum{1} (a) to (d). The original 'GI' aperture, and the reconstructed images from non-binarized samples, optimal-binarized samples and the first-$q$-binarized samples are listed in sequence. Considering the capability of reconstructing gray object, we replace the binary object aperture 'GI' by a gray object pattern - the chinese letter 'north'. We mount a neutral density attenuation filter (Daheng GCC-301021, transmission $~$50\%) right after the right half of the etched transmissive 'north' letter, which produce a gray object pattern 'north' with 3-level grayscale. The gray object pattern and experimental reconstructions under the same 20,000 measurements with different binarization operations are listed in the same sequence, shown in Fig.4, row \RNum{2} (a) to (d). Our experimental results with the gray object also shows high agreement with the conclusions in Ref.\cite{xie2019binary}.\\
\begin{figure}[htbp]
\centering
\includegraphics[width=0.8\textwidth]{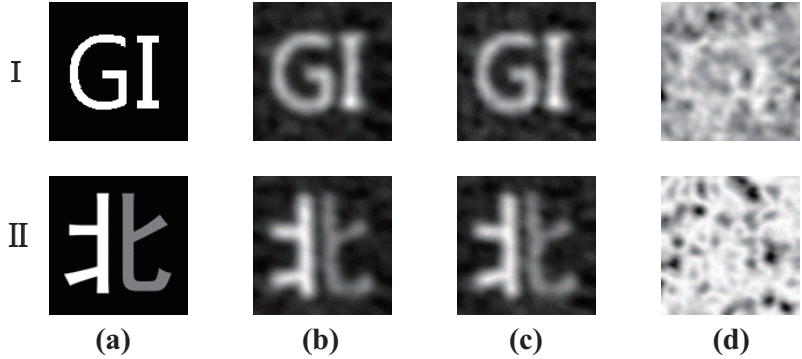}
\caption{Experimental verifications: (\RNum{1}) Binary object pattern 'GI', (\RNum{2}) Gray object pattern of chinese letter 'North'; (a)  Object aperture; Reconstructed image of 20,000 measurements with (b) no binarization, (c) optimal $q$, (d) the first $q$.}
\label{Figure4}
\end{figure}

\begin{figure}[htbp]
\centering
\includegraphics[width=1\textwidth]{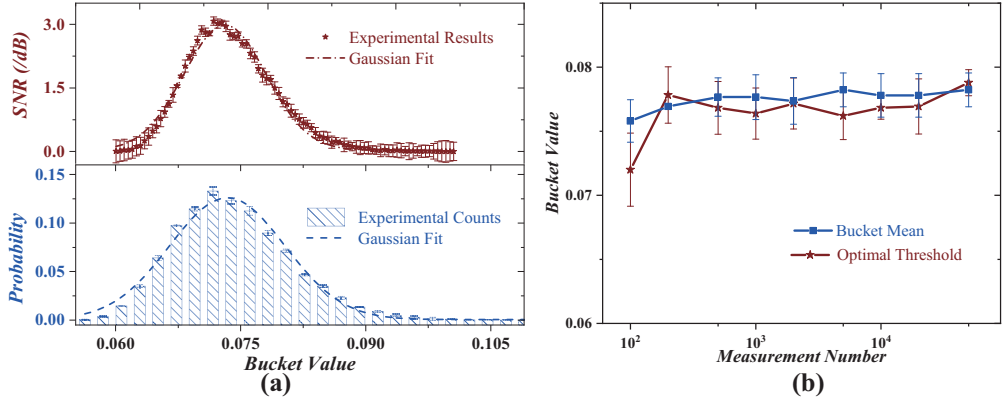}
\caption{(a) Up: Reconstructed image SNR vs. $q$ (step by 0.00069 = bucket sampling interval) of bucket values. Down: Gaussian PDF of bucket detector output values. (b) The optimal threshold and statistical mean of bucket outputs against measurement number. }
\label{Figure5}
\end{figure}
\indent We use signal-to-noise ratio (SNR) of reconstructed image to evaluate the performance of imaging system quantitively, and the definition is,
\begin{equation}
\mathrm{SNR}= 20\lg\frac{ \sum_{x} O(x) }{ \sqrt{\sum_{x}(O(x)-T(x))^{2}}},
\end{equation}
where the $O(x)$ and $T(x)$ represents the "0-1" object and reconstructed image, respectively. The higher SNR is, the better image one gets. And the SNR performance against binarization threshold $q$ under the same measurement number 20,000 is shown in upper part of Fig. 5 (a), compared with the frequency distribution of bucket outputs, both fitted well to a Gaussian PDF as assumed in our theoretical model. Apparently, there exists an optimal $q$ to approach the performance of GI without binarization, and the optimal $q$ is very close to the mean of the bucket outputs, according to the property of the Gaussian PDF. Furthermore, we make more efforts to verify if the optimal $q$ is the statistical mean of bucket outputs. So we compare the optimal $q$, which corresponds to the highest SNR of reconstructed image, and the mean of bucket outputs in Fig. 5 (b). For different measurement numbers, the optimal $q$ behaves coincidentally to the bucket mean. Here we have to mention that, the varying threshold is the physical sampling intervals of our bucket detector, which cannot be determined by the statistical properties of our bucket outputs. \\
\indent Further experimental verifications also show the performance comparison of GI with different binarization strategies from under-sampling (100) to over-sampling (50,000) conditions, the SNR of optimal binarized GI is very close to the ordinary one, which is far more better than the first-threshold binarized GI, as shown in Fig. 6 (a). For comparison, the SNR of binarized GI with the bucket mean threshold are also presented, which behaves almost the same to the optimal binarization. What's more, there still exist a gap of SNR between the convergence upper limit of optimal binarized GI and the ordinary one, and the gap has shown a clue to shrink with the increasing measurement number.\\
\begin{figure}[htbp]
\centering
\includegraphics[width=1\textwidth]{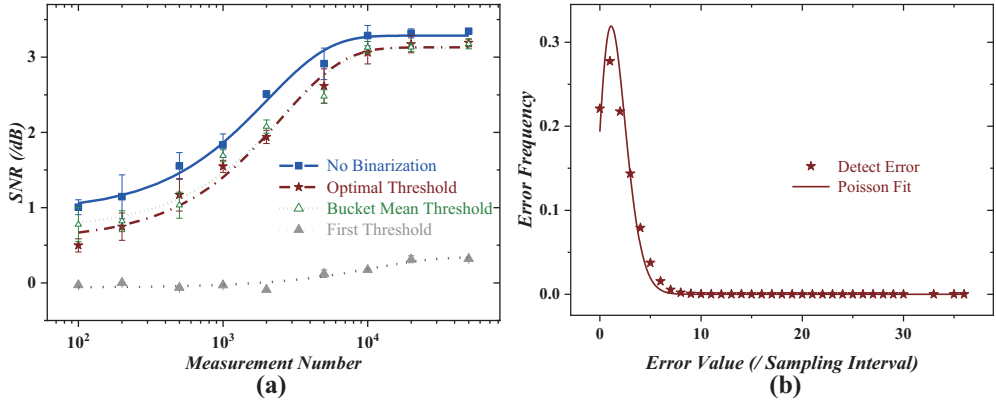}
\caption{(a) Reconstructed image SNR of the ordinary, optimal binarized and the first-threshold binarized GI against measurement number. Reconstructions with bucket mean binarization threshold are listed for comparison. (b) Detection error distribution of bucket samples under 50,000 measurements.}
\label{Fig6}
\end{figure}
\indent Here, it is noted that the behavior of the experimental results is qualitatively similar to the theoretical CRLB calculations, while we can also find some apparent discrepancies between the theoretical and the experimental results. For instance, the mismatch between the optimal $q$ and mean of bucket outputs. This stems from several possibilities, such as the mismatch of noise model, the quantization error, the non-constant of background fluctuations, polarization effects, and some unaccounted aberrations in our experiments. And we attribute this mainly to the detect error in our experiment. For estimating the noise level, we pick the bucket outputs induced by the same light fields, and record the error distribution in Fig. 6 (b), which can be modeled as a Poisson noise with an intensity of 2.2 times the sampling intervals. Due to the space constraints, we leave further discussions on this additional noise.  \\
\section{Conclusion \& Discussion}
\indent In this work we have proposed and demonstrated a general model for optimizing the performance of binary sampling GI, enabling the informative-optimal image reconstructions subject to the explicit GI system conditions. To be sure, the CRLB of binary sampling is always larger than the ordinary one. It is not surprising that binary sampling loses information, but our theoretical model indicates that the binary sampling scheme have the potential to behave arbitrarily close to the ordinary one with the optimal binarization threshold and large number of coincident measurements, which is the surprise. The optimal binarization threshold, corresponding to the CRLB minimum, are coincident with the statistical mean of bucket samples. Furthermore, we may ask the question, what would we do to optimize the performance if we can only observe binary samples? It should be noticed that the optimal binarization threshold also determines an optimal distribution of the sequence of binary signals, which is an uniform distribution on the "0-1" binary sequence, to extract the maximum information about the object. Thus, the optimization problem of binary samples can be solved by adjusting the system design of GI or the charteristics of detectors to meet the optimal distribution of binary sequence.\\
\indent For the experimental verifications based on pseudo-thermal light GI, we optimize the binarization threshold of bucket signals to maintain highest image quality assessed by SNR. The performance after optimization and the behavior of optimal binarization threshold both verify our theoretical analysis and demonstrate the effectiveness of our optimization method. Moreover, the similar behavior of the CRLB and image quality assessments, e.g., the image SNR, against the varying binarization threshold suggest that the CRLB is not only a mathematical limit, but indeed a promising candidate for optimization criterion, which yields a measureable performance benefit. \\
\indent For the future technical improvements, this optimized binary strategy is not only beneficial to the ordinary GI, since binary data can dramatically reduce sampling, storage, transfer, and calculation cost, but also pave the way for fundamentally optimizing the system design of GI in many challenging applications.\\

\section*{Fundings}
National Natural Science Foundation of China (61631014, 61401036, 61471051); National Science Fund for Distinguished Young Scholars of China (61225003); the BUPT Excellent Ph.D. Students Foundation (CX2019224).\\

\section*{Disclosures}
The authors declare that there are no conflicts of interest related to this article.\\


\bibliography{GIwithOBS}

\begin{thebibliography}{10}
\newcommand{\enquote}[1]{``#1''}

\bibitem{pittman1995optical}
T.~Pittman, Y.~Shih, D.~Strekalov, and A.~V. Sergienko, \enquote{Optical
  imaging by means of two-photon quantum entanglement,}
  {\protect\JournalTitle{Physical Review A}} \textbf{52}, R3432 (1995).

\bibitem{valencia2005two}
A.~Valencia, G.~Scarcelli, M.~D'~Angelo, and Y.~Shih, \enquote{Two-photon
  imaging with thermal light,} {\protect\JournalTitle{Physical Review Letters}}
  \textbf{94}, 063601 (2005).

\bibitem{Xue2014Lensless}
X.-F. Liu, X.-H. Chen, X.-R. Yao, W.-K. Yu, G.-J. Zhai, and L.-A. Wu,
  \enquote{Lensless ghost imaging with sunlight,} {\protect\JournalTitle{Optics
  Letters}} \textbf{39}, 2314--2317 (2014).

\bibitem{bromberg2009ghost}
Y.~Bromberg, O.~Katz, and Y.~Silberberg, \enquote{Ghost imaging with a single
  detector,} {\protect\JournalTitle{Physical Review A}} \textbf{79}, 053840
  (2009).

\bibitem{Sun2012Normalized}
B.~Sun, S.~S. Welsh, M.~P. Edgar, J.~H. Shapiro, and M.~J. Padgett,
  \enquote{Normalized ghost imaging,} {\protect\JournalTitle{Optics Express}}
  \textbf{20}, 16892--16901 (2012).

\bibitem{li_negative_2017}
J.~Li, B.~Luo, D.~Yang, L.~Yin, G.~Wu, and H.~Guo, \enquote{Negative
  exponential behavior of image mutual information for pseudo-thermal light
  ghost imaging: observation, modeling, and verification,}
  {\protect\JournalTitle{Science Bulletin}} \textbf{62}, 717--723 (2017).

\bibitem{Wang:16}
L.~Wang and S.~Zhao, \enquote{Fast reconstructed and high-quality ghost imaging
  with fast walsh-hadamard transform,} {\protect\JournalTitle{Photon. Res.}}
  \textbf{4}, 240--244 (2016).

\bibitem{katz2009compressive}
O.~Katz, Y.~Bromberg, and Y.~Silberberg, \enquote{Compressive ghost imaging,}
  {\protect\JournalTitle{Applied Physics Letters}} \textbf{95}, 131110 (2009).

\bibitem{li_image_2017}
J.~Li, D.~Yang, B.~Luo, G.~Wu, L.~Yin, and H.~Guo, \enquote{Image quality
  recovery in binary ghost imaging by adding random noise,}
  {\protect\JournalTitle{Opt. Lett.}} \textbf{42}, 1463 (2017).

\bibitem{shapiro2008computational}
J.~H. Shapiro, \enquote{Computational ghost imaging,}
  {\protect\JournalTitle{Physical Review A}} \textbf{78}, 061802 (2008).

\bibitem{duarte2008single}
M.~F. Duarte, M.~A. Davenport, D.~Takhar, J.~N. Laska, T.~Sun, K.~F. Kelly, and
  R.~G. Baraniuk, \enquote{Single-pixel imaging via compressive sampling,}
  {\protect\JournalTitle{IEEE signal processing magazine}} \textbf{25}, 83--91
  (2008).

\bibitem{liu2018fast}
X.~Liu, J.~Shi, X.~Wu, and G.~Zeng, \enquote{Fast first-photon ghost imaging,}
  {\protect\JournalTitle{Scientific Reports}} \textbf{8}, 5012 (2018).

\bibitem{li_single-photon_2019}
Z.-P. Li, X.~Huang, Y.~Cao, B.~Wang, Y.-H. Li, W.~Jin, C.~Yu, J.~Zhang,
  Q.~Zhang, C.-Z. Peng, F.~Xu, and J.-W. Pan, \enquote{Single-photon
  computational 3d imaging at 45 km,} {\protect\JournalTitle{arXiv:1904.10341
  [physics]}}  (2019). ArXiv: 1904.10341.

\bibitem{Deng:17}
C.~Deng, L.~Pan, C.~Wang, X.~Gao, W.~Gong, and S.~Han, \enquote{Performance
  analysis of ghost imaging lidar in background light environment,}
  {\protect\JournalTitle{Photon. Res.}} \textbf{5}, 431--435 (2017).

\bibitem{shi2018image}
X.~Shi, X.~Huang, S.~Nan, H.~Li, Y.~Bai, and X.~Fu, \enquote{Image quality
  enhancement in low-light-level ghost imaging using modified compressive
  sensing method,} {\protect\JournalTitle{Laser Physics Letters}} \textbf{15},
  045204 (2018).

\bibitem{ke2016fast}
J.~Ke and E.~Y. Lam, \enquote{Fast compressive measurements acquisition using
  optimized binary sensing matrices for low-light-level imaging,}
  {\protect\JournalTitle{Optics Express}} \textbf{24}, 9869--9887 (2016).

\bibitem{frieden1972restoring}
B.~R. Frieden, \enquote{Restoring with maximum likelihood and maximum entropy,}
  {\protect\JournalTitle{Journal of the Optical Society of America}}
  \textbf{62}, 511--518 (1972).

\bibitem{kay1993fundamentals}
S.~M. Kay, \emph{Fundamentals of statistical signal processing} (Prentice Hall
  PTR, 1993).

\bibitem{thiebaut2017principles}
{\'E}.~Thi{\'e}baut and J.~Young, \enquote{Principles of image reconstruction
  in optical interferometry: tutorial,} {\protect\JournalTitle{Journal of the
  Optical Society of America A}} \textbf{34}, 904--923 (2017).

\bibitem{Klyshko1988The}
D.~N. Klyshko, \enquote{The effect of focusing on photon correlation in
  parametric light scattering,} {\protect\JournalTitle{Zhurnal Eksperimentalnoi
  I Teroreticheskoi Fiziki}} \textbf{94}, 82--90 (1988).

\bibitem{Belinskii1994Two}
A.~V. Belinskii and D.~N. Klyshko, \enquote{Two-photon optics: diffraction,
  holography, and transformation of two-dimensional signals,}
  {\protect\JournalTitle{Soviet Journal of Experimental \& Theoretical
  Physics}} \textbf{78}, 259--262 (1994).

\bibitem{shih2014introduction}
Y.~Shih, \emph{An introduction to quantum optics: photon and biphoton physics}
  (Taylor \& Francis, 2014).

\bibitem{goodman2015statistical}
J.~W. Goodman, \emph{Statistical optics} (John Wiley \& Sons, 2015).

\bibitem{helstrom1969quantum}
C.~W. Helstrom, \enquote{Quantum detection and estimation theory,}
  {\protect\JournalTitle{Journal of Statistical Physics}} \textbf{1}, 231--252
  (1969).

\bibitem{yang2011bits}
F.~Yang, Y.~M. Lu, L.~Sbaiz, and M.~Vetterli, \enquote{Bits from photons:
  Oversampled image acquisition using binary poisson statistics,}
  {\protect\JournalTitle{IEEE Transactions on image processing}} \textbf{21},
  1421--1436 (2011).

\bibitem{chao_fisher_2016}
J.~Chao, E.~Sally~Ward, and R.~J. Ober, \enquote{Fisher information theory for
  parameter estimation in single molecule microscopy: tutorial,}
  {\protect\JournalTitle{J. Opt. Soc. Am. A}} \textbf{33}, B36--B57 (2016).

\bibitem{gatti2004ghost}
A.~Gatti, E.~Brambilla, M.~Bache, and L.~A. Lugiato, \enquote{Ghost imaging
  with thermal light: comparing entanglement and classicalcorrelation,}
  {\protect\JournalTitle{Physical Review Letters}} \textbf{93}, 093602 (2004).

\bibitem{Nan:17}
S.~Nan, Y.~Bai, X.~Shi, Q.~Shen, L.~Qu, H.~Li, and X.~Fu, \enquote{Experimental
  investigation of ghost imaging of reflective objects with different surface
  roughness,} {\protect\JournalTitle{Photon. Res.}} \textbf{5}, 372--376
  (2017).

\bibitem{wang_airborne_2018}
C.~Wang, X.~Mei, L.~Pan, P.~Wang, W.~Li, X.~Gao, Z.~Bo, M.~Chen, W.~Gong, and
  S.~Han, \enquote{Airborne {Near} {Infrared} {Three}-{Dimensional} {Ghost}
  {Imaging} {LiDAR} via {Sparsity} {Constraint},} {\protect\JournalTitle{Remote
  Sensing}} \textbf{10}, 732-- (2018).

\bibitem{zhao2012ghost}
C.~Zhao, W.~Gong, M.~Chen, E.~Li, H.~Wang, W.~Xu, and S.~Han, \enquote{Ghost
  imaging lidar via sparsity constraints,} {\protect\JournalTitle{Applied
  Physics Letters}} \textbf{101}, 141123 (2012).

\bibitem{Li:15}
X.~Li, C.~Deng, M.~Chen, W.~Gong, and S.~Han, \enquote{Ghost imaging for an
  axially moving target with an unknown constant speed,}
  {\protect\JournalTitle{Photon. Res.}} \textbf{3}, 153--157 (2015).

\bibitem{xie2019binary}
P.~Xie, X.~Shi, X.~Huang, Y.~Bai, and X.~Fu, \enquote{Binary detection in ghost
  imaging with preserved grayscale,} {\protect\JournalTitle{The European
  Physical Journal D}} \textbf{73}, 102 (2019).

\end{thebibliography}

\end{document}